# Boron Nitride Substrates for High Mobility Chemical Vapor Deposited Graphene

W. Gannett[1,2], W. Regan[1,2], K. Watanabe[3], T. Taniguchi[3], M. Crommie[1,2], A. Zettl[1,2,*]

[1]Department of Physics, University of California, Berkeley, California, 94720, USA

[2]Materials Sciences Division, Lawrence Berkeley National Lab, Berkeley, California, 94720, USA

[3]Advanced Material Laboratory, National Institute for Materials Science, 1-1 Namiki, Tsukuba 305-0044, Japan

*To whom correspondence should be addressed: azettl@berkeley.edu

Abstract:

Chemical vapor deposited (CVD) graphene is often presented as a scalable solution to graphene device fabrication, but to date such graphene has exhibited lower mobility than that produced by exfoliation. Using a boron nitride underlayer, we achieve mobilities as high as 37 000 cm$^2$/Vs, an order of magnitude higher than commonly reported for CVD graphene and better than most exfoliated graphene. This result demonstrates that the barrier to scalable, high mobility CVD graphene is not the growth technique but rather the choice of a substrate that minimizes carrier scattering.

Graphene has shown much promise as a material for next generation electronics due to its unique mechanical and electronic properties, including ambipolar conductivity, linear dispersion, and pseudospin.[1-6] The scalability of graphene devices has been a cause for concern, since mechanical exfoliation produces extremely low areal yields.[1] Chemical vapor deposition (CVD) growth of graphene has been demonstrated as a route to continuous monolayers many cm in width,[7] but corresponding electron mobilities are typically an order of magnitude lower than those in exfoliated graphene.[8] This reduced mobility has been attributed to higher concentrations of point defects, smaller grain sizes, and residual chemical impurities from the transfer or growth processes. Such CVD-specific scattering occurs in addition to substrate interactions and phonon effects.

One possible solution for increasing mobility in substrate-supported graphene has been to use hexagonal boron nitride (h-BN) as a layer on top of $SiO_2$. The h-BN's strong in-plane bonds, large bandgap, and planar structure provide an ideal flat, insulating, and inert surface, isolating the graphene from $SiO_2$, which has been shown to adversely affect the mobility.[9-11] The exfoliation of such h-BN flakes has been explored using Raman, optical, and transmission electron microscopy to characterize sheet thicknesses.[12-14] By exfoliating h-BN and transferring exfoliated monolayer graphene on top, electron mobilities of up to 60,000 $cm^2$/Vs have been reported.[11] Scanning tunneling spectroscopy studies have shown that fluctuations in potential and roughness of graphene on h-BN have been reduced by two orders of magnitude compared to graphene on $SiO_2$.[15,16] By comparing transport data for CVD graphene on $SiO_2$ and h-BN, we hope to learn about scattering processes specific to CVD graphene.

In this experiment, we use large high purity h-BN crystals synthesized with high pressure techniques.[17] Following prior work on exfoliated graphene and h-BN, the h-BN substrates are tape exfoliated (3M 600) onto silicon substrates with 300 nm of oxide and pre-patterned Cr/Au alignment marks.[3,12,14] These substrates are then calcined in an open-ended quartz tube in a CVD furnace at 450°C for 2 hours to remove tape residue. We grow graphene on Cu foil using a 2-step low pressure CVD process.[8,18] We then transfer graphene to the h-BN using a sacrificial PMMA layer, verify that the graphene is single layer with Raman spectroscopy, and map suitable graphene-on-BN or –oxide regions with scanning electron microscopy (SEM). One candidate region is seen in Figure 1(e). We pattern graphene with a lithographically-defined oxygen reactive ion etch and contact it with electron-beam evaporated Cr/Au electrodes (4nm/50nm). These steps are shown schematically in Figure 1 (a)-(d), and a final optical image of one device is shown in Figure 1(f). Four-probe electronic measurements are performed before and after annealing in hydrogen and argon for 3 hours at 340°C.[19]

We calculate mobilities from the slope of the conductivity versus gate voltage. Since the graphene on BN has a lower capacitance than on $SiO_2$ alone, we correct the specific capacitance using the measured thicknesses of the BN flakes obtained from atomic force microscopy, which ranged from 40 to 75 nm, assuming that the h-BN has a static dielectric constant of 4.[20] The slope is obtained from a line fit over the linear region closest to the charge neutrality point. Transport measurements at 4.2 K prior to annealing show that graphene devices on bare $SiO_2$ have electron mobilities between 4000 and 5400 $cm^2$/Vs and charge neutrality points near zero (-17 V to +2 V), while graphene devices on h-BN have electron mobilities of 1900 to 5500 $cm^2$/Vs but are consistently n-doped, with charge neutrality points from -34 V to below -50 V. Conductivity plots for representative devices on oxide and on h-BN are shown in Figure 2(a)-(b). The observed n-doping is consistent with previous work,[11,21] and the mobilities of both groups are typical for CVD graphene.[8,22] All samples also exhibit sublinear behavior at higher gate voltages, attributed to the presence of short range scattering centers in the graphene.[23]

The same measurements are performed after annealing in hydrogen and argon as specified above. Graphene devices on oxide show slightly reduced mobilities compared to the pre-anneal data (~3000 cm$^2$/Vs) and become strongly p-doped. Graphene devices on BN exhibit similar positive shifts in the charge neutrality point, moving from n-doped to essentially undoped. In addition, devices on BN all exhibit marked increases in their electron mobilities. Conductivity plots for annealed devices on oxide and on h-BN are shown in Figure 2(a)-(b). From initial data taken at 4.2 K, a simple linear fit gives electron mobility values up to 28 800 cm$^2$/Vs, a factor of 3x to 5x higher than pre-annealing and ~10x higher than their counterparts on oxide. This is 80% higher than the highest mobility for CVD-grown graphene to date.[18] This value is reproducible in-situ but does not persist through subsequent exposure to air (see below). A comparison of mobility data pre- and post-annealing is shown in Table I.

Moving beyond a simply linear fit allows us to separate the effects of different types of carrier scattering. Assuming a model of combined Coulomb ($\tau_c \sim \sqrt{n}$) and short-range ($\tau_s \sim 1/\sqrt{n}$) scattering[23] gives

$$\sigma^{-1} = (ne\mu_c + \sigma_0)^{-1} + \sigma_s^{-1} \qquad (1)$$

where $\mu_c$ is the mobility due to Coulomb scattering alone, $\sigma_0$ is the residual conductivity at the charge neutrality point, and $\sigma_s^{-1}$ is the charge density independent resistivity due to short range scattering.[23] This form fits our data adequately, allowing us to remove the effect of short range scattering and obtain the residual conductivity, the short-range scattering resistivity, and, most importantly, the mobility due to purely Coulombic scattering.[24] This method gives us electron mobilities of up to 37 000 cm$^2$/Vs prior to prolonged air exposure.

The sensitivity of SiO$_2$- and fluoropolymer-supported graphene to gas species has been previously observed, resulting in electronic doping from adsorbates.[25-27] One would expect that a sample with a clean, inert underlayer would exhibit higher sensitivity to such contaminants due to relatively less scattering from substrate interaction. To explore this effect with an h-BN underlayer, samples are exposed to air for several weeks followed by a second hydrogen anneal for 90 minutes. Transport measurements are then repeated over a range of temperatures to investigate the origin of scattering in the graphene devices.

Temperature series data appear in Figure 3(a). The data are taken as the temperature increases from 4.2 K to 293 K with the exception of an initial room temperature measurement prior to cooling. Other than the 293 K and 4.2 K measurements, the mobilities are relatively insensitive to temperature; the slight downward slope suggests a component from electron-phonon scattering. The drop between 4.2 K and 6 K may be due to contaminants driven from the sample heater and adsorbed on the cold sample, which is unable to recover until the material desorbs above 170 K. (The heater is not used at 4.2 K, and the mobility is therefore preserved.) Despite this contamination, the electron mobilities on BN are still ~3-4x higher than those on oxide.

We also examine the conductivity minima of our samples, shown in Figure 3(b). They show higher values at room temperature and 4 K, consistent with increased scattering from cryosorbed species. Most values fall between 6 and $8e^2/h$, typical of graphene samples without strong intervalley scattering.[28] There appears to be no significant difference between the conductivity minima of graphene on oxide or on h-BN.

In summary, we demonstrate high-mobility devices from CVD graphene on exfoliated h-BN and, in the process, show that known CVD techniques are more than adequate for producing substrate-supported devices consistently above 10 000 cm$^2$/Vs. Our electron mobility of 28 800 cm$^2$/Vs is 80% higher than the highest reported value for CVD graphene and many times higher than commonly reported values.[8,18,22,29] The discovery of a more scalable production method for suitable substrates would allow the production of wafer scale high mobility (>20 000cm$^2$/Vs) graphene devices, a key goal since graphene was first isolated.

Acknowledgements:
The authors thank J. H. Chen for helpful discussion and B. Alemán for help with AFM. This research was supported in part by the U. S. National Science Foundation under Grant No. 0906539 which provided for experiment design, sample fabrication, and transport characterization. Support was also received from the Director, Office of Energy Research, Materials Sciences and Engineering Division, of the U. S. Department of Energy under Contract No. DE-AC02-05CH11231 through the sp$^2$-bonded Materials Program which provided for Raman and SEM characterization, and the Office of Naval Research MURI program under Grant No. N00014-09-1-1066 which provided for graphene synthesis. W.R. acknowledges support through a National Science Foundation Graduate Research Fellowship.

Table I: Electron mobilities at 4.2 K based on linear fits.

| Substrate | $\mu_e$, pre-anneal (cm$^2$/Vs) | $\mu_e$, post-anneal (cm$^2$/Vs) | $\mu_e$, re-annealed (cm$^2$/Vs) |
|---|---|---|---|
| Oxide | 4000-5400 | 3000-3300 | 1900-2300 |
| h-BN | 1900-5500 | 9200-28 800 | 10 000-14 000 |

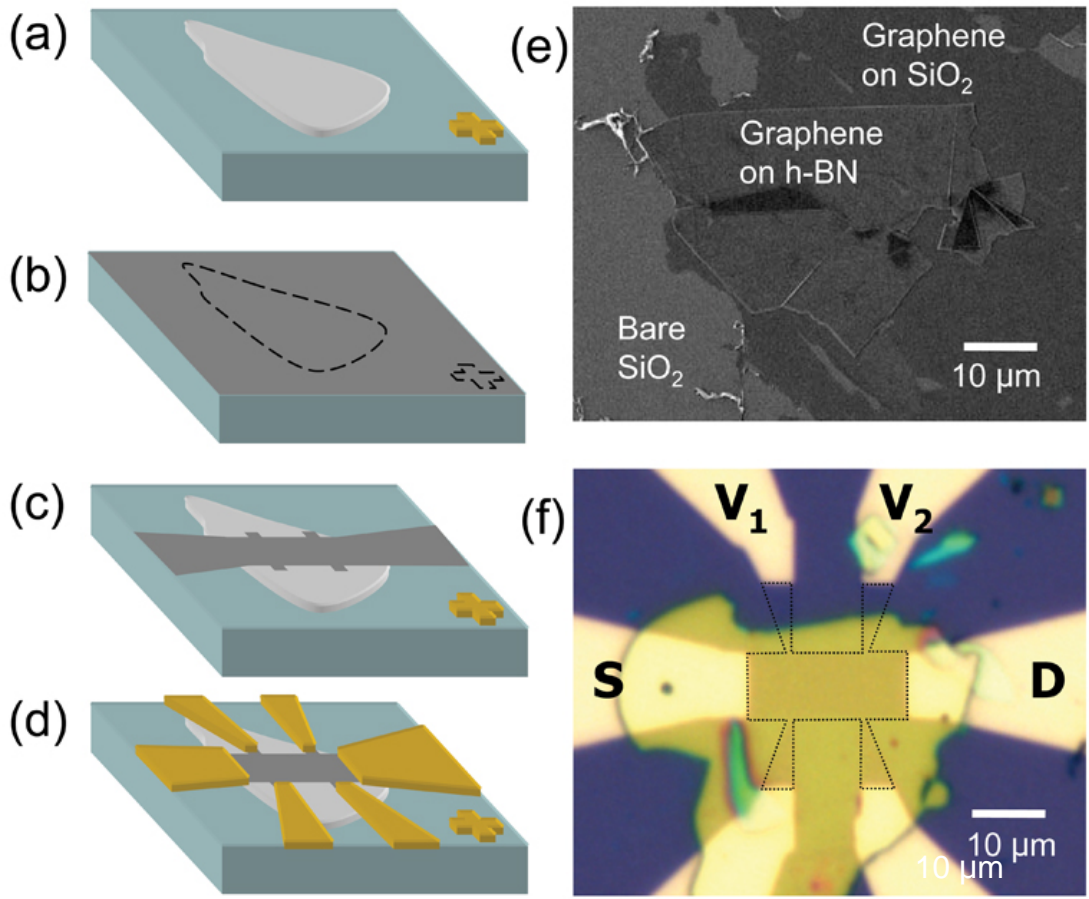

Figure 1: (a)-(d) sample fabrication steps shown schematically. a) h-BN is exfoliated onto a $SiO_2$ substrate with existing alignment marks. b) CVD-grown graphene is deposited over the entire chip. c) Graphene is patterned with e-beam lithography and reactive ion etching. d) Electrodes are fabricated with e-beam lithography and e-beam evaporation of Cr and Au. e) SEM image of graphene on a candidate h-BN flake. f) Optical image of finished graphene-on-BN device, with electrodes labeled and graphene marked with a dotted outline.

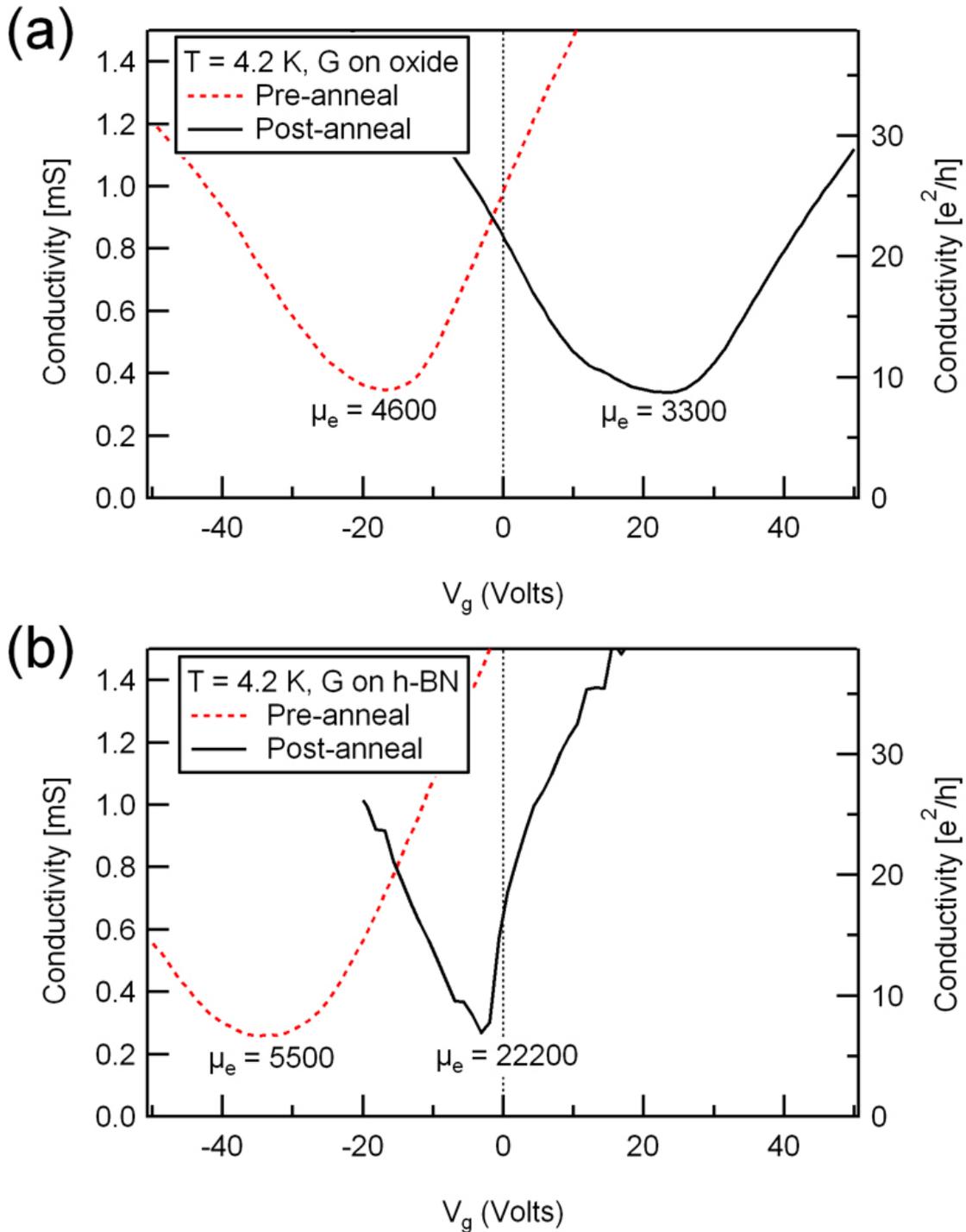

Figure 2: (a)-(b) Conductance versus gate voltage for CVD graphene before and after annealing on $SiO_2$ and h-BN, respectively. All four sweeps are from positive to negative gate voltage.

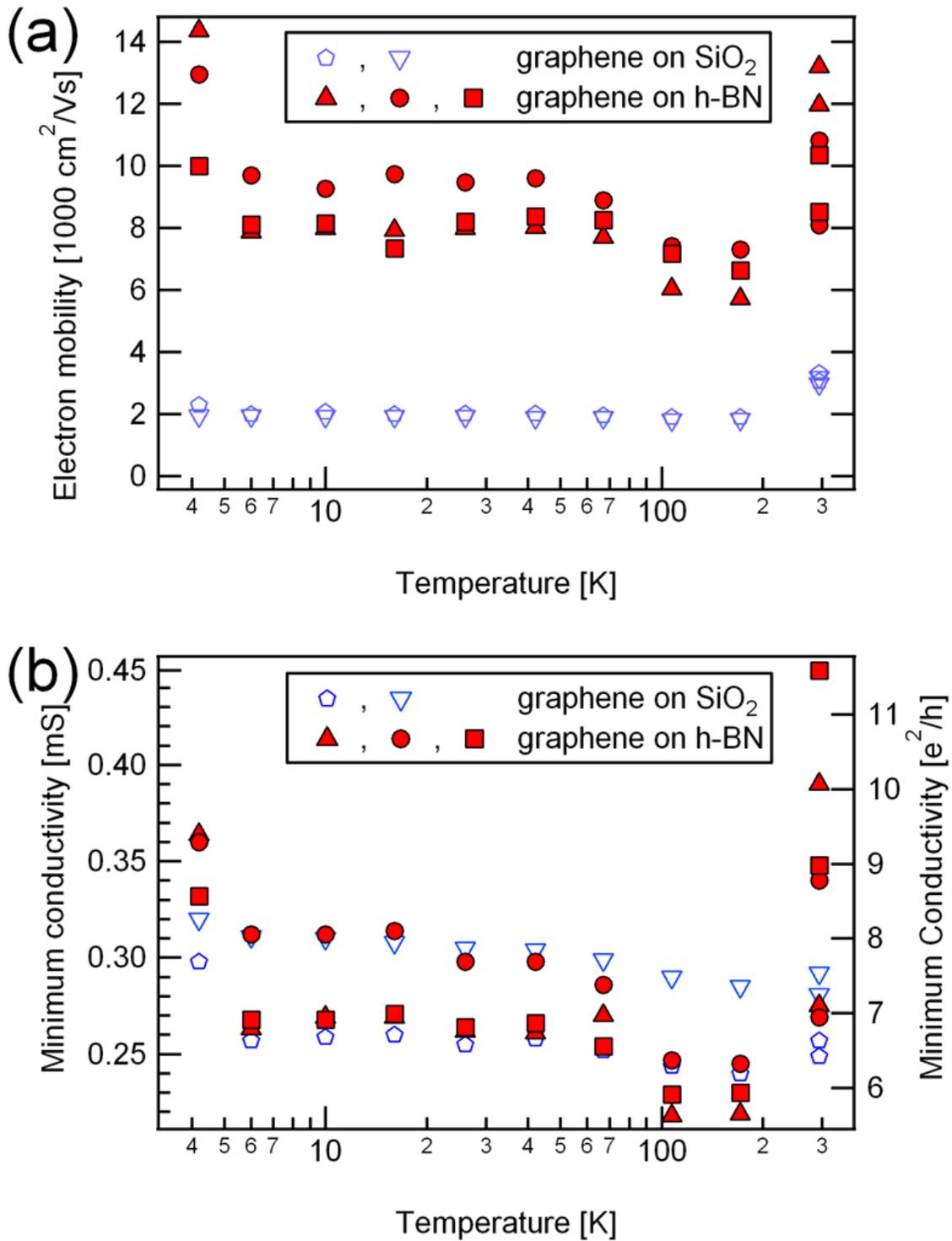

Figure 3: a) Electron mobility versus temperature for 2 graphene-on-oxide and 3 graphene-on-BN samples. The oxide samples have consistently lower mobilities across all temperatures. b) Conductivity minima versus temperature for the same devices, vertical axis is given in both mS and units of $e^2/h$.